\documentclass[showpacs,twocolumn,aps]{revtex4}
\usepackage{amssymb}
\usepackage{amsmath}
\usepackage{graphicx}
\usepackage{lscape}
\usepackage{booktabs}
\begin{document}
\title{Centrality dependence of light (anti)nuclei and (anti)hypertriton
production\\
in Au+Au collisions at $\sqrt{s_{\rm{NN}}}$ = 200 GeV}
\author{Gang Chen\footnote{Corresponding Author: chengang1@cug.edu.cn}, Huan Chen, Juan Wu, De-Sheng Li and Mei-Juan Wang}
\address{
 School of Mathematics and Physics, China University of Geoscience, Wuhan
430074, China.}

\begin{abstract}
We have used the dynamically constrained phase space coalescence
model to investigate the centrality dependence of light (anti)nuclei
and (anti)hypertriton production based on the $6.2\times 10^7$
hadronic final states generated by the PACIAE model in Au+Au
collisions at $\sqrt{s_{\rm{NN}}}=200$ GeV in $|y| <1$ and $p_T<5$
acceptances. It turned out that the yields of light (anti)nuclei and
(anti)hypertriton strongly depend on the centrality, i.e. their
yields decrease rapidly with the increase of centrality bins; but
their yield ratios are independent on centrality. These theoretical
results are consistent with the STAR and PHENIX data. Furthermore,
centrality distribution of $d$ ($\overline d$), $^3{He}$
($^3{\overline{He}}$) and $_{\overline\Lambda}^3H$
($\overline{_{\overline\Lambda}^3H}$) follows Gaussian
distributions. This means that light (anti)nuclei and
(anti)hypertriton are primarily produced in the central collisions.

\end{abstract}

\maketitle

\section{Introduction}
Ordinary matter and antimatter asymmetry is a fundamental problem in
modern physics research. Since C. D. Anderson discovered the first
antiparticle, i.e. positron, inside the cosmic rays in 1932, then
the anti-neutrons, anti-protons and other anti-particles and light
(anti)nuclei have gradually been discovered in scientific
experiments~\cite{1956,1965,1970,1974,prl2005,star2007}. It is
believed that there exists equal amount of ordinary matter and
antimatter in the initial stages of the universe. Especially, after
the discovery of the first hypernuclei in 1952~\cite{1953},
searching for (anti)hypernuclei bound states and exploring the
hyperon-nucleon interaction have been steadily fascinating the
sights of nuclear physicists~\cite{star2,star3}. Because an
anti-nucleus is very unstable and its yield is very low, so the
progress of antimatter research was slow. In the high energy
collision experiments, the high temperature and high baryon density
matter 
is similar to the "Fireball" environment produced in the initial
stages of the Big Bang, which is uniquely suitable for the
production of both the light (anti)nuclei and (anti)hypernuclei
mentioned above.

The STAR collaboration has reported their measurements of
$^3_{\Lambda}H$ , $\overline{_{\overline\Lambda}^3H}$ and
$\overline{^4He}$ in Au+Au collisions at the top RHIC energy
~\cite{star2,star3}. The ALICE collaboration has also published
their preliminary $\overline d$ yield of $\sim6\times 10^{-5}$
measured in the pp collisions at $\sqrt s=7$~TeV
~\cite{alice,alice1}.

On the other hand, the theoretical study of light nuclei
(anti-nuclei) is usually divided into two steps. Firstly the
nucleons and hyperons are calculated with some selected models, such
as the transport models. Then the light nuclei (anti-nuclei) are
calculated by the phase space coalescence model~\cite{grei,chen,ma}
and/or the statistical model~\cite{pop,peter} etc. Recently, the
production of light nuclei (hypernuclei) in the Au+Au/Pb+Pb
collisions at relativistic energies has been investigated
theoretically by the coalescence+blast-wave method~\cite{ma1} and
the UrQMD-hydro hybrid model+thermal model~\cite{stei},
respectively.

Besides, we have proposed an approach, PACIAE + the dynamically
constrained phase-space coalescence model ({\footnotesize DCPC}
model)~\cite{yuyl}. {\footnotesize DCPC} is based on the final
hadronic state generated by a parton and hadron cascade model
PACIAE~\cite{sa2}. We have first predict the light nuclei
(anti-nuclei) yield, transverse momentum distribution, and the
rapidity distribution in non-single diffractive pp collisions at
$\sqrt{s }=7$~TeV~\cite{yuyl}. Then we use this method to
investigate the light nuclei (anti-nuclei) and hypernuclei
(anti-hypernuclei) productions for in 0-5\% most central Au+Au
collisions at $\sqrt{s_{\rm{NN}}}=200$~GeV~\cite{Cheng}.

It should be noted that the results of light (anti)nuclei and
(anti)hypernuclei yields and their ratios given by different
experimental groups employed event samples with different
centralities. For example, the STAR analysis in ref.~\cite{0909}
used a data sample of 25 million central triggered events (0-12\%
centrality) plus 24 million minimum-bias triggered events (0-80\%
centrality). For STAR in ref.~\cite{star2}, about 89 million
collision events were collected using minimum-bias events, and an
additional 22 million events were collected using near-zero impact
parameter collisions. The data set for PHENIX analysis in
ref.~\cite{prl2005} includes $21.6\times 10^6$ minimum bias events.
In theoretical studies, the production of light (anti)nuclei and
(anti)hypernuclei is usually investigated in most central heavy ion
collisions events (0-5\% centrality)~\cite{ma1,stei,yuyl,Cheng}.
However, in previous theoretical and experimental studies, little
attention has been paid to the impact of the centrality on the
production for light (anti)nuclei and (anti)hypernuclei in heavy ion
collisions. In fact, the centrality in heavy ion collisions may have
a great impact on the yields of light (anti)nuclei. In order to
effectively compare the experimental and theoretical results for
different collision centralities, it is necessary to study the
dependence of light (anti)nuclei and (anti)hypernuclei yields as
well as their ratios on the centrality in detail.

The paper is organized as follows: In Sec.~II, we briefly introduce
the PACIAE model and the dynamically constrained phase-space
coalescence model (DCPC model). In Sec.~III, the numerical results
of light (anti)nuclei and (anti)hypernuclei yields and their ratios
are given in different centrality bins, and are compared with the
STAR and PHENIX data. A short summary is the content of Sec. IV.

\section {MODELS}
The PYTHIA model (PYTHIA 6.4~\cite{sjo2}) is devised for the high
energy hadron-hadron ($hh$) collisions. In this model, a hh
collision is decomposed into parton-parton collisions. The hard
parton-parton scattering is described by the leading order
perturbative QCD (LO-pQCD) parton-parton interactions with a
modification of parton distribution function in a hadron. For the
soft parton-parton collision, a non-perturbative process is
considered empirically. The initial- and final-state QCD radiations
and the multiparton interactions are also taken into account.
Therefore, the consequence of a hh collision is a partonic multijet
state composed of di-quarks (anti-diquarks), quarks (antiquarks) and
gluons, as well as a few hadronic remnants. This is then followed by
the string construction and fragmentation. A hadronic final state is
obtained for a hh collision eventually.

The parton and hadron cascade model PACIAE~\cite{sa2} is based on
PYTHIA 6.4 and is devised mainly for the nucleus-nucleus collisions.
In the PACIAE model, firstly, the nucleus-nucleus collision is
decomposed into the nucleon-nucleon (NN) collisions according to the
collision geometry and NN total cross section. Each NN collision is
described by the PYTHIA model with the string fragmentation
switches-off and di-quarks (anti-diquarks) randomly breaks into
quarks (anti-quarks). So the consequence of a NN collision is a
partonic initial state composed of quarks, anti-quarks, and gluons.
Provided all NN collisions are exhausted, one obtains a partonic
initial state for a nucleus-nucleus collision. This partonic initial
state is regarded as the quark-gluon matter (QGM) formed in the
relativistic nucleus-nucleus collisions. Secondly, the parton
rescattering proceeds. The rescattering among partons in QGM is
randomly considered by the 2$\rightarrow$ 2 LO-pQCD parton-parton
cross sections~\cite{comb}. In addition, a $K$ factor is introduced
here to account for higher order and non-perturbative corrections.
Thirdly, the hadronization happens after the parton rescattering.
The partonic matter can be hadronized by the Lund string
fragmentation regime~\cite{sjo2} and/or the phenomenological
coalescence model~\cite{sa2}. Finally, the hadronic matter proceeds
rescattering until the hadronic freeze-out (the exhaustion of the
hadron-hadron collision pairs). We refer to~\cite{sa2} for the
details.

In quantum statistical mechanics~\cite{kubo} one can not precisely
define both position $\vec q\equiv (x,y,z)$ and momentum $\vec
p\equiv (p_x,p_y,p_z)$ of a particle in the six dimension phase
space, because of the uncertainty principle
\begin{equation*}
\Delta\vec q\Delta\vec p \leqslant h^3.
\end{equation*}
We can only say that this particle lies somewhere within a six
dimension quantum "box" or "state" with a volume of $\Delta\vec
q\Delta\vec p$. A particle state occupies a volume of $h^3$ in the
six dimension phase space~\cite{kubo}. Therefore one can estimate
the yield of a single particle by defining the following integral

\begin{equation}
Y_1=\int_{H\leqslant E} \frac{d\vec qd\vec p}{h^3},
\end{equation}
where $H$ and $E$ are the Hamiltonian and energy of the particle,
respectively. Similarly, the yield of N particle cluster can be
estimated as following integral
\begin{equation}
Y_N=\int ...\int_{H\leqslant E} \frac{d\vec q_1d\vec p_1...d\vec
q_Nd\vec p_N}{h^{3N}}. \label{phas}
\end{equation}

Therefore the yield of $\overline{_{\overline\Lambda}^3H}$ in the
dynamically constrained phase space coalescence model, for instance,
is assumed to be
\begin{align}
Y_{\overline{_{\overline\Lambda}^3H}}=\int ...
\int\delta_{123}\frac{d\vec q_1d\vec p_1
  d\vec q_2d\vec p_2d\vec q_3d\vec p_3}{h^{9}},
\label{yield}
\end{align}
where
\begin{align}
 \delta_{123}=\left\{
  \begin{array}{ll}
  1 \hspace{0.2cm} \textrm{if} \hspace{0.2cm} 1\equiv \bar p, 2\equiv \bar n,
    3\equiv \bar\Lambda,\\
    \hspace{0.75cm} m_0\leqslant m_{inv}\leqslant m_0+\Delta m,\\
    \hspace{0.75cm}  q_{12}\leqslant D_0, \hspace{0.2cm} q_{13}
    \leqslant D_0, \hspace{0.2cm} q_{23}\leqslant D_0, \\
  0 \hspace{0.2cm}\textrm{otherwise},
  \end{array}
  \right.
\label{yield1}
\end{align}
\begin{equation}
m_{inv}=[(E_1+E_2+E_3)^2-(\vec p_1+\vec p_2+\vec p_3)^2]^{1/2},
\label{yield2}
\end{equation}
and ($E_1,E_2,E_3$) and ($\vec p_1,\vec p_2,\vec p_3$) are the
energies and momenta of particles $\bar p,\bar n,\bar\Lambda$,
respectively. In Eq.~(\ref{yield1}), $m_0$ and $D_0$ stand for,
respectively, the rest mass and diameter of
$\overline{_{\overline\Lambda}^3H}$, $\Delta m$ refers to the
allowed mass uncertainty, and $ q_{ij}=|\vec q_{i}-\vec q_{j}|$ is
the vector distance between particle $i$ and $j$.

 As the hadron position and momentum distributions
from transport model simulations are discrete, the integral over
continuous distributions in Eq.~(\ref{yield}) should be replaced by
the sum over discrete distributions. In a single event of the final
hadronic state obtained from transport model simulation, the
configuration of $\overline{_{\overline\Lambda}^3H}$ ($\bar p$+$\bar
n$+$\bar\Lambda$) system, for instance, can be expressed as
\begin{equation}
C_{\bar p\bar n\bar\Lambda}(\Delta q_1,\Delta q_2,\Delta q_3;\vec
p_1,\vec p_2,\vec p_3), \label{conf}
\end{equation}
where the subscripts $1\equiv \bar p$, $2\equiv \bar n$, $3\equiv
\bar \Lambda$, and $\Delta q_i$ refers to the distance between
particle $i$ and center-of-mass of three particles, ie.
\begin{equation} \Delta q_i = |\vec q_i - \vec q_c |, \ \ \ (i=1,2,3).\end{equation}
Here, $\vec q_c$ is the coordinate vector of the center-of-mass of
$\bar p$, $\bar n$, and $\bar\Lambda$. Therefore the third
constraint (diameter constraint) in Eq.~(\ref{yield1}) is
correspondingly replaced by
\begin{equation}
\Delta q_1\leqslant R_0, \ \ \ \hspace{0.2cm} \Delta q_2\leqslant
R_0, \ \ \ \hspace{0.2cm}\Delta q_3\leqslant R_0,
\end{equation}
where $R_0$ refers to the radius of
$\overline{_{\overline\Lambda}^3H}$.

Each of the above configurations contributes a partial yield of
\begin{align}
y_{\bar p\bar n\bar\Lambda}=&\left\{
  \begin{array}{ll}
  1 \hspace{0.2cm} \textrm{if} \hspace{0.2cm} m_0\leqslant m_{inv}\leqslant
    m_0+\Delta m,\\
    \hspace{0.4cm} \Delta q_1\leqslant R_0, \hspace{0.2cm} \Delta q_2\leqslant R_0,
    \hspace{0.2cm} \Delta q_3\leqslant R_0; \\
  0 \hspace{0.2cm}\textrm{otherwise};
  \end{array}
  \right.
\label{yield3}
\end{align}
to the yield of $\overline{_{\overline\Lambda}^3H}$. So the total
yield of $\overline{_{\overline\Lambda}^3H}$ in a single event is
the sum of the above partial yield over the configuration in
Eq.~(\ref{conf}) and their combinations. An average yield for all
the events is required at the end.

\section{The results and discusses}
Firstly we produce the final state particles using the PACIAE model.
In the PACIAE simulations we assume that hyperons heavier than
$\Lambda$ decay already. The model parameters are fixed on the
default values given in PYTHIA~\cite{sjo2}. However, the K factor as
well as the parameters parj(1), parj(2), and parj(3), relevant to
the strange production in PYTHIA~\cite{sjo2}, are given by fitting
the STAR data of $\Lambda$, $\overline \Lambda$, $\Xi^-$, and
$\overline{\Xi^-}$ in Au+Au collisions at $\sqrt{s_
{\rm{NN}}}=200$~GeV~\cite{star5}. The fitted parameters of $K$=3
(default value is 1 or 1.5~\cite{sjo2}), parj(1)=0.12 (0.1),
parj(2)=0.55 (0.3), and parj(3)=0.65 (0.4) are used to generate
$6.2\times 10^7$ events (final hadronic states) by the PACIAE model.
A minimum-bias events sample is formed in Au+Au collisions at
$\sqrt{s_{\rm{NN}}}=200$~GeV of $|y|<1$ and $0 < p_t<5$ acceptances.
\begin{figure}[htbp]
\includegraphics[width=0.34\textwidth]{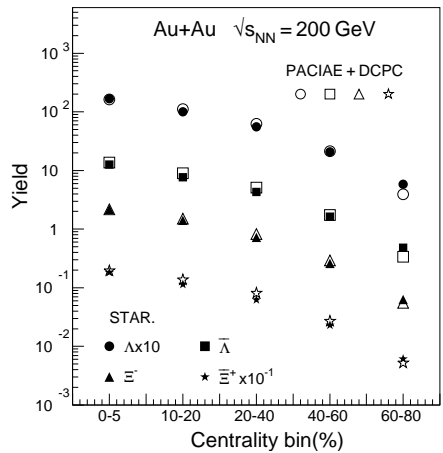}
\caption{Yields of strange particle at midrapidity ($|y|<1$ for
${\Lambda}$ and $\overline{\Lambda}$, $|y|<0.75$ for $\Xi^-$ and
$\overline{\Xi^-}$) in the Au+Au collisions at
$\sqrt{s_{\rm{NN}}}=200$~GeV  as a function of centrality bin. The
open symbols represent our model results and the solid symbols are
the data points from STAR~\cite{star5}.} \label{tu1}
\end{figure}

Fig.~\ref{tu1} shows the strange particle yields (open symbols)
calculated with the PACIAE model at the midrapidity for different
centralities Au+Au collisions at $\sqrt{s_{\rm{NN}}}=200$~GeV. The
solid symbols in this figure are the experimental data taken
from~\cite{star5}. One sees in this figure that the PACIAE results
agree well with the experimental data.

Then, the yields of $d$ ($\overline d$), $^3{He}$
($^3{\overline{He}}$), as well as $_{\overline\Lambda}^3H$
($\overline{_{\overline\Lambda}^3H}$) and their ratios are
calculated by the {\footnotesize DCPC} model for different
centrality bins of 0-5\%, 0-12\%, 0-30\%, 0-50\%, 0-80\% and
0-100\%. The subsamples of collision events in different centrality
bins are identified by the impact parameter $b$ in PACIAE model from
the minimum-bias event sample.

\begin{figure}[htbp]
\includegraphics[width=0.34\textwidth]{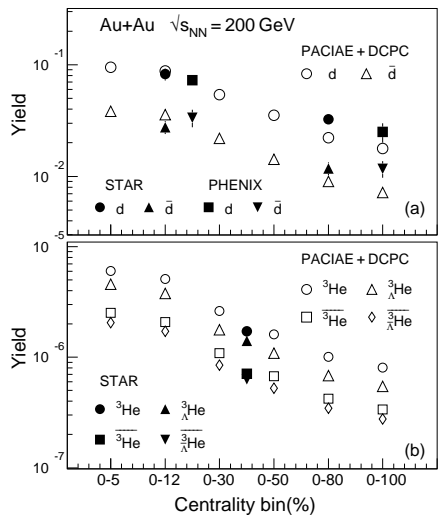}
\caption{The comparison of yields for light (anti)nuclei and
(anti)hypertriton between model results and experimental data at
midrapidity in the Au+Au collisions at $\sqrt{s_{\rm{NN}}}=200$~GeV,
plotted as a function of centrality bin. The open symbols represent
our model results. The solid symbols are the data points from
STAR~\cite{0909,star2} and PHENIX~\cite{prl2005}. Panel (a) is for
$\overline d$ and $d$, and panel (b) is for
$\overline{_{\overline\Lambda}^3 H}$, ${_{\Lambda}^3 H}$,
$^3{\overline{He}}$ and $^3{{He}}$.} \label{tu2}
\end{figure}

In Fig.~\ref{tu2} we show the yield distributions of $d$, $\overline
d$, ${_{\Lambda}^3 H}$, $\overline{_{\overline\Lambda}^3 H}$,
$^3{{He}}$, and $\overline{^3{He}}$ in different centralities Au+Au
collisions at $\sqrt{s_{\rm{NN}}}=200$~GeV. The upper panel is for
$d$ ($\overline d$), the lower panel is for ${_{\Lambda}^3 H}$
($\overline{_{\overline\Lambda}^3 H}$) and $^3{{He}}$
($\overline{^3{He}}$). For comparison, the figure also exhibits the
experimental data with solid symbols. It can be seen from
Fig.~\ref{tu2} that the yields of light (anti)nuclei and
(anti)hypertriton all decrease rapidly with the centrality bins; the
{\footnotesize PACIAE+DCPC} model results (the open symbols) are in
agreement with the STAR~\cite{0909,star2} and PHENIX~\cite{prl2005}
data(the solid symbols).

\begin{table*}[htbp]
\caption{The yield ratios in different centrality bins from Au+Au
collisions at $\sqrt{s_{\rm{NN}}}=200$~GeV,  comparing with PHNEX
and STAR data.}
\begin{tabular}{cccccccccccc}
\hline  \hline
 Ratio&
\multicolumn{6}{c}{$R_\textrm{PACIAE+DCPC}^a$}&\multicolumn{3}{c}{$R_\textrm{STAR}$}
&\multicolumn{2}{c}{$R_\textrm{PHNEX}$} \\ \cline{2-12}
Centrality(\%) \ \ \
 &0-5&0-12&0-30&0-50&0-80&0-100&\ \ \ \ 0-12&0-80&mixed&\ \ \ \ 0-20&0-100 \\ \hline
$\overline{d}/d$&0.402&0.406&0.406&0.405&0.405&0.405&\ \ \ \ 0.394$^b$&0.428$^b$&$-$&\ \ \ \ 0.462$^d$&0.468$^d$\\
${\overline{^3 He}}/^3He$ &0.418&0.407&0.416&0.417&0.419&0.419&$-$&$-$&0.45$\pm0.18\pm0.07^c$&$-$&$-$  \\
$\overline{_{\overline\Lambda}^3 H}/_{\Lambda}^3 H$ &0.451&0.455&0.480&0.481&0.507&0.507&$-$&$-$&0.49$\pm0.18\pm0.07$$^c$& $-$&$-$\\
$_{\Lambda}^3 H/^3{He}$ &0.756&0.739&0.674&0.677&0.677&0.677&$-$&$-$&0.82$\pm0.16\pm0.32$$^c$& $-$&$-$\\
${\overline{_{\overline\Lambda}^3 H}/^3\overline{He}}$&0.817&0.826&0.778&0.782&0.819&0.819 &$-$&$-$&0.89$\pm0.28\pm0.13$$^c$&$-$&$-$ \\
\hline \hline
\multicolumn{12}{l}{$^a$ calculated with $\Delta m$=0.0003 GeV for $d, \overline{d}$ and $\Delta m$=0.0002 GeV for $\overline{^3 He}, ^3He, \overline{_{\overline\Lambda}^3 H}$ and $_{\Lambda}^3 H$.} \\
\multicolumn{12}{l}{$^b$ taken from Fig.~\ref{tu2} in~\cite{0909}.} \\
\multicolumn{12}{l}{$^c$ taken from~\cite{star2} calculated with 89
million minimum-bias events and 22 million central collision
 events.}\\
\multicolumn{12}{l}{$^d$ taken from~\cite{prl2005}.}
\\\end{tabular} \label{paci1}
\end{table*}

In Tab.~\ref{paci1} the yield ratios of light antinuclei and
antihypertriton to light nuclei and hypertriton, as well as
$^3_{\Lambda}H$ to $^3He$ and $\overline{^3_{\Lambda}H}$ to
$\overline {^3{He}}$ are given in different centralities Au+Au
collisions at $\sqrt{s_{\rm{NN}}}=200$~GeV. For comparison,
experimental results from STAR and PHENIX are also given in
Tab.~\ref{paci1}. One can see in this table that the yield ratios in
different centrality bins remain unchanged although their yields
decrease rapidly with the centrality bin as shown Fig.~\ref{tu2},
and the results obtained from our model are in agreement with the
experimental data from STAR~\cite{0909,star2} and
PHENIX~\cite{prl2005}.

\begin{figure}[htbp]
\includegraphics[width=0.38\textwidth]{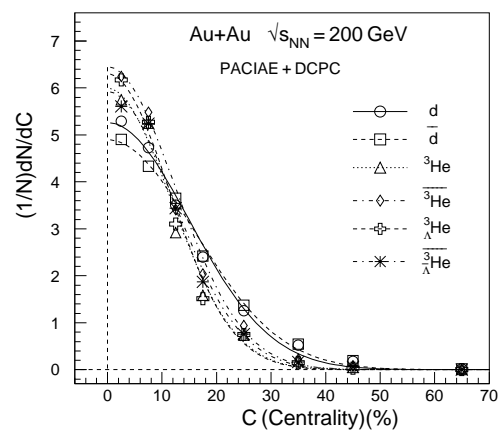}
\caption{Centrality distribution of particle yields $dN/dC/N$ for
light (anti)nuclei and (anti)hypertriton ($d$, $\overline d$,
$_{\Lambda} ^3H$, $\overline{_{\overline\Lambda}^3 H}$, $^3{He}$ and
$^3{\overline{He}}$) at midrapidity in Au+Au collisions at
$\sqrt{s_{\rm{NN}}}=200$~GeV. The results are obtained by the
{\footnotesize PACIAE+DCPC} model. The curves are fitted by the
Gaussian distribution Eq.(10)} \label{tu3}
\end{figure}

In order to further analyze the centrality dependence of light
(anti)nuclei generation, the centrality distributions of particle
yield for $d$, $\overline d$, ${_{\Lambda}^3 H}$,
$\overline{_{\overline\Lambda}^3 H}$, $^3{{He}}$ and
$\overline{^3{He}}$ in Au+Au collisions at $\sqrt{s_{\rm{NN}}}=
200$~GeV, calculated by the {\footnotesize PACIAE+DCPC} model, are
given in Fig.~\ref{tu3}. It shows that the centrality distributions
of particle yield are similar for six kinds of particles. Their
yields have a maximum peak nearby centrality $C = 0$, and fall
rapidly with the increase of centrality until they approach zero at
centralities $C$ greater than about 40\%. This indicates that the
yield of the light (anti)nuclei and (anti)hypertriton is strongly
dependent on the centrality, i.e. the light (anti)nuclei and
(anti)hypertriton are generated primarily in central collision
region, and their yields in peripherad collision are almost
negligible.

Quantitatively, the results are in accord with the Gaussian
distributions, i.e.
\begin{equation}
\frac{1}{N}\frac{dN}{dC}=\alpha{\rm exp}(-\frac{C^2}{2\sigma^2}).
\label{conf1}
\end{equation}
The fitted parameters $\alpha$ and $\sigma$ corresponding to
different particles are shown in Tab.~\ref{paci2}. We can further
obtain that the value of $3\sigma$ is ($44.7\pm1.2$)\% for $d$,
($47.7\pm1.2$)\% for $\overline d$, $(35.1\pm1.2$)\% for $^3He$,
$(37.2\pm0.9$) for ${\overline{^3 He}}$, $(34.2\pm1.2$) for
$_{\Lambda}^3 H$ and  $(36.6\pm0.9$) for
${\overline{_{\overline\Lambda}^3 H}}$.

\begin{table*}[htbp]
\caption{The parameters $\alpha$ and $\sigma$  fitted by Eq.(10)
with Fig.~\ref{tu3}.}
\begin{tabular}{ccccccc}
\hline  \hline
par.
&d &$\overline{d}$ &$^3He$ &${\overline{^3 He}}$ &$_{\Lambda}^3 H$&${\overline{_{\overline\Lambda}^3 H}}$ \\ \hline
$\alpha$&$5.26\pm0.15$&$4.85\pm0.15$&$6.00\pm0.17$&$6.36\pm0.16$&$6.32\pm0.17$&$5.92\pm0.16$\\
$\sigma$ &$0.149\pm0.004$&$0.159\pm0.004$&$0.117\pm0.004$&$0.124\pm0.003$&$0.114\pm0.004$&$0.122\pm0.003$ \\
\hline \hline
\end{tabular} \label{paci2}
\end{table*}

\begin{figure}[htbp]
\includegraphics[width=0.4\textwidth]{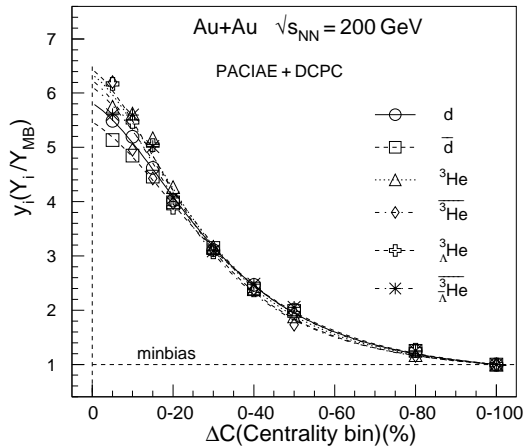}
\caption{ The normalized yield $y_{i}$ for the $\overline d$, $d$,
$\overline{_{\overline\Lambda}^3 H}$, ${_{\Lambda}^3 H}$,
$^3{\overline{He}}$ and $^3{{He}}$ as a function of centrality bin.
The results are calculated by {\footnotesize PACIAE+DCPC} model at
midrapidity in the Au+Au collisions at $\sqrt{s_{\rm{NN}}}=200$~GeV.
The curves are fitted by Eq.(12).}
 \label{tu4}
\end{figure}

In order to effectively compare the theoretical and experimental
results for different centrality bins, we define a normalized yield
 $y_{i}$ as
\begin{equation}
y_{i}\equiv \frac{Y_i}{Y_{\rm{MB}}},
\end{equation}
where $Y_i$ is the yield in the $i$-th centrality bin, $Y_{\rm{MB}}$
is the yield in minimum-bias events. The normalized yield $y_{i}$
for $\overline d$, $d$, $\overline{_{\overline\Lambda}^3 H}$,
${_{\Lambda}^3 H}$, $^3{\overline{He}}$ and $^3{{He}}$ in Au+Au
collisions at $\sqrt{s_{\rm{NN}}}=200$~GeV are given in
Fig.~\ref{tu4}. In this figure the curse is a function of
\begin{equation}
y_{i}=\frac{Y_i}{Y_{\rm{MB}}}= \gamma{\frac{1-\Delta C}{\beta +
\Delta C^2}} +1,
\end{equation}
fit to the data points. In the equation, $\gamma$ and $\beta$ are
parameters. It can be seen in Fig.~\ref{tu4} that the normalized
yields $y_{i}$ for light (anti)nuclei and (anti)hypertriton also
decrease rapidly with the increase of centrality bins until they
approach 1 at $\Delta C=1$, i.e. $y_i =Y_{MB}\equiv 1$.

Setting the $y_{i}$ distribution as a reference calibration curve,
the yields in any centrality bins can be translated into the
constant yield $Y_{MB}$ corresponding to its minimum-bias events
according to Eq.(11). Thus theoretic and experimental results in
different centralities heavy ion collisions can be directly
compared.

\section{Conclusion}
In this paper we use {\footnotesize PACIAE+DCPC} model to
investigate the centrality dependence of light (anti)nuclei and
(anti)hypertriton production in Au+Au collisions at top RHIC energy.
The results show that the yields of $d$, $\overline d$,
$\overline{_{\overline\Lambda}^3 H}$, ${_{\Lambda}^3 H}$,
$\overline{^3{He}}$, and $^3{{He}}$ decrease rapidly with the
increase of centrality bins. However, the yield ratios of
$\overline{D}$ to $D$, $\overline{_{\overline\Lambda}^ 3H}$ to
${_{\Lambda}^3 H}$ and $^3{\overline{He}}$ to $^3{{He}}$, as well as
$\overline{_{\overline\Lambda}^ 3H}$ to $^3{\overline{He}}$ and
${_{\Lambda}^3 H}$ to $^3{{He}}$ are independent on centrality. The
results obtained from our model are also consistent with the STAR
and PHENIX data. Our models results also show that the light
(anti)nuclei and (anti)hypertriton are primarily produced in central
collision regions, and the dependence of the yields on centrality
follows a Gaussian distributions with centrality $C < 45$\% for $d$
($\overline d$), $C < 35$\% for $^3{He}$ ($^3{\overline{He}}$) and
$_{\overline\Lambda}^3H$ ($\overline{_{\overline\Lambda}^3H}$). The
yields in the peripherad collision events are almost negligible. By
defining a normalized yield $y_{i}$, the theoretical and
experimental results of yields in different centrality bins of heavy
ion collisions can be converted into the constant yield $Y_{MB}$ to
be directly compared.

\begin{center} {ACKNOWLEDGMENT} \end{center}
Finally, we acknowledge the financial support from 
Central Universities (GUGL 100237,120829,130249) in China. The
authors thank Prof. Ben-Hao Sa and PH.D. YU-liang Yan  for helpful
discussions.

\end{document}